\documentclass[prb,floatfix,showpacs,twocolumn,superscriptaddress]{revtex4}
\usepackage{graphics}
\usepackage{graphicx}
\usepackage{dcolumn} 
\usepackage{bm}
\usepackage{float}
\usepackage{epsfig}
\begin{document}
\def\rhov{{\mbox{\boldmath{$\rho$}}}}
\def\tauv{{\mbox{\boldmath{$\tau$}}}}
\def\Deltav{{\mbox{\boldmath{$\Delta$}}}}
\def\Lambdav{{\mbox{\boldmath{$\Lambda$}}}}
\def\Thetav{{\mbox{\boldmath{$\Theta$}}}}
\def\Psiv{{\mbox{\boldmath{$\Psi$}}}}
\def\Phiv{{\mbox{\boldmath{$\Phi$}}}}
\def\epsilonv{{\mbox{\boldmath{$\epsilon$}}}}
\def\sigmav{{\mbox{\boldmath{$\sigma$}}}}
\def\alphav{{\mbox{\boldmath{$\alpha$}}}}
\def\xiv{{\mbox{\boldmath{$\xi$}}}}
\def\oh{{\scriptsize 1 \over \scriptsize 2}}
\def\ot{{\scriptsize 1 \over \scriptsize 3}}
\def\of{{\scriptsize 1 \over \scriptsize 4}}
\def\tf{{\scriptsize 3 \over \scriptsize 4}}

\title{Symmetry Analysis of Multiferroic Co$_3$TeO$_6$}

\author{A. B. Harris$^\dagger$}

\affiliation{Department of Physics and Astronomy, University of
Pennsylvania, Philadelphia PA 19104}
\date{\today}
\begin{abstract}
A phenomenological explanation of the magnetoelectric behavior
of Co$_3$TeO$_6$ is developed.  We explain the second harmonic
generation data and the magnetic field induced spontaneous
electric polarization in the magnetically ordered phase below
20K.
\end{abstract}
\pacs{61.50.Ks,61.66.-f,63.20.-e,76.50.+g}
\maketitle

\section{INTRODUCTION} 

Recently there has been an explosion in the number of compounds
which exhibit nontrivial magnetoelectric behavior at low
temperatures.[\onlinecite{NATURE,TMOPRL,NVOPRL,NVOPRB,ABHPRB}]
Co$_3$TeO$_6$ (CTO) is an interesting such system whose properties
have recently been studied.[\onlinecite{HUDL}]  Although the
magnetic structure is as yet not clarified, it seems useful
to construct a mean-field scenario which can explain
the major experimental results.  The measurments of Hudl 
{\it et al.}[\onlinecite{HUDL}] of $M/H$, $d(M/H)/dT$, and
$C/T$ versus $T$, where $M$ is the magnetization, $H$ the magnetic
field, and $C$ the specific heat,
indicate that there are at least two magnetic phase transitions
at temperatures below about 30K, one at $T_1 \approx 26$K and
another at $T_2\approx 18.5$K, but the details of the magnetic 
tructure are not known, other than that the system 
is not ferromagnetic.  According to
Ref. [\onlinecite{IVAN}], the magnetic structure is
described by several incommensurate wave vectors.  Single crystal
netron diffraction measurements reveal that the incooomensurate
wave vector(s) are in the a-b plane and not along c.[\onlinecite{LI}]
We propose the existence of magnetic order at zero wave vector,
consistent with the results of Li {\it et al.},
although,  this would have to  involve an
antiferromagnetic arrangement of moments within the unit cell
to give the observed zero net moment. In addition, our analysis 
suggests the appearance of an additional magnetic phase transition.
In the absence of magnetic order the crystal symmetry
is[\onlinecite{BECKER,HUDL}] that of space group C2/c
(\#15 in Ref. \onlinecite{ITC}).  We will take the generators
of this space group to be the glide operation $m_b \equiv (x, -y,z+1/2)$,
a two-fold screw rotation about the crystal $b$ axis, 
$2_b \equiv (-x,y+1/2,-z+1/2)$, and the three translations,
$(x+1/2,y+1/2,z)$,  $(x-1/2,y+1/2,z)$, and  $(x,y,z+1)$, where
$x$, $y$, and $z$ are in units of lattice constants.
These are equivalent to those of Ref. \onlinecite{HATCH}.

\section{EXPERIMENTAL DATA}

The data of Hudl {\it et al.}[\onlinecite{HUDL}]
consist of several types.  As mentioned
above, the measurements of magnetization and specific heat
indicate phase transitions at at least two temperatures, $T_1$
and $T_2$, but the nature of magnetic ordering could not be determined
from their data.  The lower-temperature transition may
be a discontinuous one. Of primary interest to us is their
measurement of the intensity of second harmonic generation (SHG),
whose cross section is proportional to the third order electric
susceptibility $\chi_{\alpha \beta \gamma}$, where $\alpha$, $\beta$
and $\gamma$ label components (or in the present case label
crystallographic directions). Their experimental geometries 
are chosen such that the SHG cross section is proportional to
$\chi_{\alpha \alpha \alpha}$.  In a system having high symmetry,
{\it e. g.} having inversion symmetry, the SHG intensity is zero for 
all frequencies, and this applies to CTO above about $T_2=18.5$K. 
However, below that temperature they find that
$\chi_{aaa}$ and $\chi_{ccc}$ are nonzero, but
$\chi_{bbb}$ is apparently zero at all temperatures.
From this they conclude that the point group retains only $m_b$
symmetry. As we shall see, if, as they assert, the symmetry is
magnetically broken, this is not a correct conclusion.

Another type of data of crucial interest to us the
measurement of the electric polarization, $\bf P$ in the $a$-$c$
plane, as a function of temperature and 
magnetic field for magnetic fields along the
crystallographic $a$ and $c$ directions.  For zero magnetic field,
at temperatures below about 18K, they find a very small, possibly
zero, spontaneous polarization in the $a$ and $c$ directions
which increases almost proportional to the magnetic field.
In fact, we find that their results for $P_c$ at $T=5$K as a
function of $H_a$ can be fit within experimental uncertainty 
($\pm 5$ in $P_c$) to
\begin{eqnarray}
P_c &=& -0.15 + 6.93H_a + 0.33H_a^2 \ ,
\end{eqnarray}
with $P_c$ in $\mu$C/m$^2$ and $H$ in Tesla.
In other words, they found an important magnetic field-dependent 
contribution to $P_c$ linear in $H_a$ with $P_c(H_a=0)\approx 0$.

\section{SYMMETRY ANALYSIS}

We will carry out our analysis in terms of an expansion
about the ``vacuum'', which we take to be the phase above
26K in which the magnetic order parameters and electric 
polarization are zero. Magnetically ordered phases
are described by nonzero magnetic order parameters.
We will also discuss briefly nonmagnetic structural distortions
which lower the crystal symmetry from C2/c and which
are described by appropriate order parameters.
Although, as mentioned in Ref. \onlinecite{HUDL}, there
may exist incommensurate magnetic order described by ${\bf M}({\bf q})$
with $q \not= 0$,  incommensurate magentic order can not,
by itself, explain the experimental results, as we will explain below.

\subsection{Electric Polarization}

We first review the phenomenological theory of magnetization induced
electric polarization ${\bf P}$.  The magnetoelectric
free energy is of the form
\begin{eqnarray}
F_{\rm ME} &=& \frac{1}{2} \chi_E^{-1} {\bf P}^2 + \sum_n
\Delta F^{(n)} \ ,
\end{eqnarray}
where $\chi_E$ is the dielectric susceptibility of the vaccum
(the phase above $T=26$K), which we assume to be isotropic for
simplicity  and $\Delta F^{(n)}$ is the contribution  linear in
${\bf P}$ (so that it induces a nonzero value of ${\bf P}$)
and of order $H^n$. For instance, to lowest order in powers
of the magnetic order parameters, we write[\onlinecite{ABHPRB,FIEBIG}]
\begin{eqnarray}
\Delta F^{(0)} &=& \sum_{{\bf q} \not= 0}
a_{\alpha kl} ({\bf q}) P_\alpha [M_k({\bf q})^*M_l({\bf q})
- M_k({\bf q})M_l({\bf q})^*] \nonumber \\
&& + b_{\alpha kl} P_\alpha M_k(q=0) M_l(q=0) \nonumber \\
\Delta F^{(1)} &=& c_{\alpha \beta k} P_\alpha H_\beta M_k(q=0)
\nonumber \\ 
\Delta F^{(2)} &=&  \sum_{\bf q \not= 0}
d_{\alpha \beta \gamma kl} ({\bf q}) P_\alpha H_\beta H_\gamma
\nonumber \\ && \times
[M_k({\bf q})^*M_l({\bf q}) - M_k({\bf q})M_l({\bf q})^*] \nonumber \\
&& + e_{\alpha \beta \gamma kl} P_\alpha H_\beta H_\gamma
M_k(q=0) M_l(q=0) \ ,
\end{eqnarray}
where we invoke the Einstein convention which implies
summation over repeated subscripts,
Greek subscripts label crystallographic directions, and Roman letters
label irreducible representations (irreps), which in the present case are
one dimensional. The magnetic order parameter $M_k({\bf q})$ can be
thought of as the amplitude of the magnetic normal mode associated with 
irrep $\Gamma_k$.[\onlinecite{ABHPRB}]  These normal modes are the
linear combinations of magnetic moments within the unit which bring the
quadratic terms in the Landau expansion into diagonal form. We will discuss
the symmetry of the $M_k$'s in a moment.  Here the Fourier transforms are
defined so that for ${\bf q} \not= 0$, ${\cal I}M_k({\bf q})=M_k({\bf q})^*$,
where ${\cal I}=m_b2_b$ is spatial inversion.

$F_{\rm ME}$ must be invariant under all the symmetries of the
``vacuum."  These symmetries include time reversal symmetry, 
translatational symmetry (which leads to wave vector conservation),
and the crystallographic symmetries $m_b$ and $2_b$ (which together
imply invariance under spatial inversion ${\cal I}$).  We will
consider the crystallogrphic symmetries in a moment.  Time reversal symmetry
requires that the total number of powers of $H$ and $M({\bf q})$ must
be even. The condition that $F_{ME}$ be real valued implies that
${\bf a}({\bf q})$ and ${\bf d}({\bf q})$ be pure imaginary.
The form of $\Delta F^{(1)}$ is such that wave vector conservation
implies that the magnetic order for this mechanism must occur
at zero wave vector, and, as previously noted, it must be
antiferromagnetic to be consistent with the observed zero net
magnetic moment of the system.  (In fact CTO has
a large enough paramagnetic unit cell that antiferromagnetic
order can develop without increasing the size of the unit cell, as
occurs in LaTiO$_3$[\onlinecite{SCHMITZ}] and 
Cr$_2$O$_3$[\onlinecite{SPALDIN}].) Such an antiferromagnetic moment
would be consistent with the magnetic measurements of Hudl 
{\it et al.}[\onlinecite{HUDL}]

When $F_{ME}$ is minimized with respect to ${\bf P}$ to
obtain its equilibrium value, one sees that $\Delta F^{(n)}$ gives rise 
to a contribution to ${\bf P}$ which is of order $H^n$.  In many
multiferroics, such as Ni$_3$V$_2$O$_8$[\onlinecite{NVOPRL,NVOPRB}]
(NVO) and TbMnO$_3$[\onlinecite{TMOPRL}] (TMO), $\Delta F^{(0)}$
is a crucial term which gives rise to a spontaneous polarization
at $H=0$.  Many other cases are similarly
analyzed in Ref. \onlinecite{ABHPRB}. In these cases, the magnetic 
order is incommensurate, so that the polarization (a zero wave vector
property) can not be linear in the magnetic order parameter.
Since in CTO $P \propto H$, we consider $\Delta F^{(1)}$ from which 
we get
\begin{eqnarray}
P_\alpha &=& \chi_E \sum_{\beta k} c_{\alpha \beta k}
H_\beta M_k \ .
\end{eqnarray}

We now show how the crystallographic symmetries constrain the
coefficient tensor $c_{\alpha \beta k}$. In particular, we
will show that these symmetries fix the symmetry of $M_k$.
For this purpose, note that $\Delta F^{(1)}$ has to be invariant
under these symmetries.  In this analysis, we will
confine ${\bf P}$ and ${\bf H}$ to be perpendicular to the
crystallographic ${\bf b}$ direction, as they were in the
experiments of Ref. \onlinecite{HUDL}. In that case, 
we only consider terms in $\Delta F^{(1)}$ with
$\alpha$ and $\beta$ labeling the crystallographic $a$ and $c$
directions, and $k$ labels the possible magnetic irreps
at zero wave vector.
Remembering that ${\bf H}$ is a pseudovector, we note that
\begin{eqnarray}
m_b [P_\alpha H_\beta] &=& - P_\alpha H_\beta \ , \hspace{0.4 in}
2_b [P_\alpha H_\beta] = P_\alpha H_\beta \ .
\end{eqnarray}
Accordingly, for $\Delta F^{(1)}$ to be an invariant we require that
\begin{eqnarray}
m_b M_k = - M_k \ , \hspace{0.4 in}
2_b M_k =  M_k \ .
\label{MSYM} \end{eqnarray}
To implement Eq. (\ref{MSYM}) we need to characterize the symmetry of
the magnetic ordering, which we have inferred occurs at zero wave vector.
For phase transitions the catalog of broken symmetry phases that can
result from a phase transition in any of the 230 crystallographic 
space groups can be obtained using the suite of computer programs
ISODISTORT which is accessible on the web.[\onlinecite{WEB}]
As applied to CTO one predicts that only four magnetic irreps
can result from a single phase transition at zero wave vector.
This formulation specifically does not
allow for a multicritical point at which there is a simultaneous
breaking of two distinct symmetries.  For CTO there is no
experimental indication that the magnetic phase transitions
arise from such a multicritical point.[\onlinecite{TRIC}] Therefore 
we assume the validity of the four possible magnetic phases
of Table \ref{BIBLE} which ISODISTORT lists for space group C2/c.
Looking at Table \ref{BIBLE} we see that to be consistent with
Eq. (\ref{MSYM}), the magnetic order parameter can only be that of
irrep $\Gamma_2$.

\begin{table} [h!]
\caption{\label{BIBLE} Symmetry of the magnetic irreps $\Gamma_n$ 
at zero wave vector for CTO.
Here $\lambda({\cal O})$ is the eigenvalue of the operator ${\cal O}$:
${\cal O}M_k = \lambda ({\cal O}) M_k$, where $M_k=M(\Gamma_k)$ is
the order parameter associated with the $k$th irrep.
Also ${\cal E}$ is the identity
and ${\cal I}$ is spatial inversion.  In the last line, we give the
direction of the ferromagnetic moment if it is allowed to be nonzero.}
\vspace{0.2 in}
\begin{tabular} {|| c| c| c |c|c||}
\hline  \hline 
& $\Gamma_1$ & $\Gamma_2$ & $\Gamma_3$ & $\Gamma_4$ \\
\hline
$\lambda({\cal E})$ & $+1$ & $+1$ & $+1$ & $+1$ \\
$\lambda(2_b)$ & $+1$ & $+1$ & $-1$ & $-1$ \\
$\lambda(m_b)$ & $+1$ & $-1$ & $-1$ & $+1$ \\
$\lambda({\cal I})$ & $+1$ & $-1$ & $+1$ & $-1$ \\
\hline
${\vec M}$ & $\vec b$ & $0$ & $\perp b$ & $0$ \\
\hline \hline \end{tabular}
\end{table}

\subsection{Second Harmonic Generation}

We now turn to the analysis of the SHG cross section at
$H=0$.  To develop a nonzero SHG cross section a quantity like
$\partial \chi_{\alpha \alpha \alpha}/\partial M_\beta$
must be nonzero in the vacuum (magnetically disordered phase),
so that when we turn on the magnetic order parameter
$M_\beta$ (in the magnetically ordered phase) the SHG cross
section becomes nonzero.  To study this quantity it is
useful to note that it has the symmetry of
$\partial [p_\alpha p_\alpha p_\alpha] /
\partial M_k$, where $p_\alpha$ is the $\alpha$-component of
the dipole moment operator and $M_k$ is a magnetic order
parameter.  One sees that
this quantity is zero because $M_k$ is odd under time reversal,
and the dipole moment operator is even under time
reversal.[\onlinecite{KANE}] Therefore, the phenomenological
explanation for a nonzero SHG cross section must come from
$X_\alpha \equiv \partial^2 [p_\alpha p_\alpha p_\alpha ] /
[\partial M_k({\bf q}) \partial M_l^*({\bf q})]$ being nonzero in the
disordered phase.  This quantity has the same symmetry as
$X_\alpha \equiv p_\alpha^3 {\cal M}$ where
${\cal M}=M_k({\bf q}) M_l^*({\bf q})$ or ${\cal M}=M_kM_l$.
The fact that the SHG is proportional
to the product of two different order parameters, each of which,
as we shall see, describes a one dimensional
irrep, has been noted before[\onlinecite{SHG}].
Here, from the polarization data, we know of the existence of at
least one irrep at zero wave vector and according to 
Ref. \onlinecite{IVAN} magnetic ordering occurs with
at least one irrep at nonzero wave vector.  To have a nonzero SHG
cross section we need a second irrep, either at zero wave vector
or at the same nonzero wave vector. In either case the appearance of a second
irrep requires an as yet unobserved phase transition, which may be
unobtrusive enough that it was not seen by Hudl {\it et al.}.
We consider these two scenarios in turn.

The condition for a nonzero SHG cross section
is identical to that for a nonzero electric polarization
because the symmetry properties of the dipole moment operator and
the electric polarization are the same.  Thus, if $\chi_{aaa}$ and
$\chi_{ccc}$ are nonzero, then $P_a$ and $P_c$ are expected to be
nonzero.  Furthermore, no matter which scenario
is adopted, there is a possible problem in that although 
experiments show that for $H=0$, $\chi_{aaa}$ and $\chi_{ccc}$ are
nonzero and $\chi_{bbb}=0$, the expected field
independent contributions to $P_a$ and $P_c$ are very small.
The explanation for this may be that the SHG is anomalously
large when the polarization is due to modification of electronic
orbits (as contrasted to being due to ionic
displacements).[\onlinecite{LOTTER}]

In the first scenario, we assume that the nonzero SHG cross section
is induced by magnetic order at zero wave vector and study
the symmetry properties of $X_\alpha$.  Since
$p_\alpha^2$ transforms like unity, it suffices to study
$X_\alpha \equiv p_\alpha M_kM_l$, to indicate 
whether $\chi_{\alpha \alpha \alpha}$ is or is not zero. Since
$\chi_{bbb}=0$, we require that $p_bM_kM_l$ be odd under either
$m_b$ or $2_b$.  This implies that $M_kM_l$ either be even under
$m_b$ or odd under $2_b$.  Using Table \ref{TWO},
we see that this criterion excludes either $M_1 M_2$
or $M_3 M_4$ being nonzero. Similarly if $\chi_{aaa}$
and $\chi_{ccc}$ are nonzero, we require that  both
$p_aM_kM_l$ and $p_cM_kM_l$ be even under both $m_b$ and $2_b$.
This implies that $M_kM_l$ be even under $m_b$ and odd under $2_b$.
These requirements indicate that either $M_1M_4$
or $M_2M_3$ be nonzero.  Since we have previously
invoked the existence of irrep $M_2$ to
explain the electric polarization, we opt for $M_2 M_3$
being nonzero. The fact that the magnetic moment perpendicular to
$b$ (coming from irrep $M_3$) is zero (or very small) would
have to be a result specific to the details of the interactions.

In the second scenario one would have to posit an additional
phase transition involving a second incommensurate magnetic irrep
to give rise to a nonzero SHG cross section.  In principle,
one would have an accompanying field independent polarization
coming from  $\Delta F^{(0)}$, whose absence in experiment would
have to be explained as above in terms an unusually large SHG
cross section.  To illustrate
this mechanism consider the hypothetical case when the incommensurate
magnetic ordering occurs at ${\bf q}=q_0 \hat b$.  In this case
one finds that there are two magnetic irreps, one of which, call it
$M_1({\bf q})$, is even under $2_b$ and the other, call it
$M_2({\bf q})$, is odd under $2_b$. Then one sees that
$X \equiv p_a[M_1({\bf q})^*M_2({\bf q})- M_1({\bf q})M_2({\bf q})^*]$
and
$Y\equiv p_c[M_1({\bf q})^*M_2({\bf q})- M_1({\bf q})M_2({\bf q})^*]$
are both invariant under $2_b$ (and under ${\cal I}$), so that
$\chi_{aaa} \propto X$ and $\chi_{ccc} \propto Y$ are allowed to be
nonzero, whereas $\chi_{bbb}$ remains zero. In a common 
scenario[\onlinecite{ABHPRB}] one irrep would give rise to
nonzero magnetic moments along the $\hat b$ axis, and the other
would give rise to nonzero magnetic moments along the $c$ axis.
These irreps would be out of phase (so that
$M_1({\bf q})^* M_2({\bf q}) - M_1({\bf q}) M_2({\bf q})^*$
is nonzero) giving rise to a magnetic spiral.[\onlinecite{MOST}]

\begin{table} [h!]
\caption{\label{TWO} As Table \ref{BIBLE}.  Symmetry of the
product of two zero wave vector magnetic irreps $\Gamma_n$ for CTO.}
\vspace{0.2 in}
\begin{tabular} {|| c| c| c |c|c|c|c||}
\hline  \hline 
& $\Gamma_1\Gamma_2$ & $\Gamma_1\Gamma_3$ & $\Gamma_1\Gamma_4$ &
$\Gamma_2 \Gamma_3$ & $\Gamma_2\Gamma_4$ & $\Gamma_3\Gamma_4$ \\
\hline
$\lambda(2_b)$ & $+1$ & $-1$ & $-1$ & $-1$ & $-1$ & $+1$ \\
$\lambda(m_b)$ & $-1$ & $-1$ & $+1$ & $+1$ & $-1$ & $-1$ \\
\hline \hline \end{tabular}
\end{table}

\subsection{Discussion}

To summarize our conclusions: we require the existence of
zero wavevector magnetism according to irrep $M_2$
to explain the magnetic field induced electric
polarization.  In one scenario we explain the SHG
cross section as being proportional to $M_2M_3$.
Since we prefer not to assume a multicritical point,  the latter
result would imply that there are actually two phase transitions.
At the higher-temperature transition (at $T=18.5$K) a magnetic field
induced spontaneous electric polarization appears and at the
lower-temperature transition (at some temperature close to but
below 18.5K) the SHG cross section becomes nonzero. Here
a very small magnetic field independent polarization
should also appear.  In principle, one would hope
to show the temperature dependence of the SHG cross section
to be proportional to the product of these two order parameters
whose temperature dependence was independently established by
neutron diffraction.  This type of experimental program
was carried out for the electric polarization of
NVO (see Fig. 6 of Ref. \onlinecite{NVOTWO}).
Note also a magnetically induced SHG cross section
implies that the symmetry involves
time reversal.  The magnetic phase with irrep
$M_2$ is {\it odd} under $m_b$, as indicated 
in Table \ref{BIBLE}. In contrast, if we were dealing with a
nonmagnetic structural phase transition, as the analysis of
Hudl {\it et al.} tacitly assumes, then the low-temperature
phase would be even under $m_b$, as they state. However,
note that the presence of magnetic irreps $M_2$ and $M_3$
breaks the mirror symmetry of $m_b$, but the symmetry of
$m_b$ plus time reversal is maintained.  This is consistent
with the results of Tables 7 and 4 of Ref.  \onlinecite{BIRSS}.
(The misidentification of Ref. \onlinecite{HUDL} is not
completely harmless.  If one assumes that $m_b$ symmetry is unbroken,
then, as they find, it is impossible to use $\Delta F^{(1)}$ to explain why
$\partial P_\alpha / \partial H_\beta$ is nonzero for $\alpha,\beta=a,c$.)

The second scenario has similar ramifications except that
it involves magnetic ordering at some incommensurate wave vector.
This scenario would also require a second phase transition
at which a second incommensurate order parameter would 
appear.[\onlinecite{FN}]
In principle, such a transition could involve a slightly different
wave vector than that already present.  But, as argued in Ref.
\onlinecite{NVOPRB}, quartic terms in the Landau free energy
would favor locking these two nearby wave vectors to the same value.

We have implicitly assumed that
the experimental results are induced by magnetic ordering.
One might question whether the results of Ref. \onlinecite{HUDL}
could be explained by simply invoking one or more phase
transitions driven by structural distortions. Since magnetic
ordering appears at these transitions the question is which
order parameter is the primary one whose presence induces
the appearance of the other one.  If $Q$ is a structural
order parameter (like the tilting angle of a cage of
oxygen ions), then one can invoke an interaction of the
type $V \sim M(\Gamma_k) M(\Gamma_l)Q$ to explain the
appearance of a nonzero value of $Q$ at the transition.  
Via this coupling the appearance of one or more magnetic 
order parameters (which are the primary order parameters)
would induce a structural distortion (because $Q$ appears linearly).
The converse case, where the magnetic order parameter appears
linearly and the primary order parameter $Q$ appears
quadratically (or linearly, for that matter) is not allowed
by time reversal symmetry.  But if the magnetic order parameters
are the primary ones, then the theoretical approach of the
present paper is essentially unchanged by the appearance of 
secondary structural order parameters.

\noindent {\bf ACKNOWLEDGEMENT}
I acknowledge helpful advice on magnetic
symmetry from J. Kikkawa and I thank J. Lynn for interesting me
in this system. I also gratefully acknowledge support from NIST.


\begin{thebibliography} {99}
\bibitem{TRIC} 
The phase diagram in the $H$-$T$ plane, part of which is shown
in Fig. 5 of Ref.
\onlinecite{HUDL}, could indeed have a multicritical point
at a special value of $H$.  But except near that
special point, the phase transition can be analyzed as in
Ref.  \onlinecite{WEB} or in the present paper.
\bibitem{FN} 
A single incommensurate irrep usually describes a collinear modulated
magnetic structure.  As the temperature is lowered, one expects that
a second irrep may appear, giving rise to a magnetic spiral for which
the fixed spin length constraint is 
satisfied.[\onlinecite{KAPLAN,SP1,SP2}]
\end{thebibliography}
\end{document}